\documentclass[12pt]{article}
\usepackage{amsmath}
\usepackage{amssymb}
\usepackage{dsfont}
\usepackage{cite}
\newsymbol \blackbox 1004
\newcommand{\eh}{\hfill}\newlength{\sperr}

\newenvironment{proof}{{\settowidth{\sperr}{\bf\rm
Proof}%
\par\addvspace{0.3cm}\noindent\parbox[t]{1.3\sperr}
{\bf\rm P\eh r\eh o\eh o\eh f\eh }%
}}{\nopagebreak\mbox{}
$\blackbox$\par\addvspace{0.3cm}}

\def\nn{\nonumber}

\def\Lam{\Lambda}
\def\s{\sigma}

\def\wt{\widetilde}
\def\ov{\overline}

\def\p{\partial}

\def\BC{{\mathbb C}}
\def\BR{{\mathbb R}}
\def\BN{{\mathbb N}}

\def\cla{{\mathcal A}}

\def\clx{{\mathcal X}}

\def\cli{{\mathcal I}}

\def\im{{\rm Im\ }}

\newcommand{\E}{\mathrm{e}}
\newcommand{\I}{\mathrm{i}}

\newtheorem{Pa}{Paper}[section]
\newtheorem{Tm}[Pa]{{\bf Theorem}}

\newtheorem{Cy}[Pa]{{\bf Corollary}}
\newtheorem{Rk}[Pa]{{\bf Remark}}

\newtheorem{Ee}[Pa]{{\bf Example}}

\newtheorem{Pn}[Pa]{{\bf Proposition}}

\title{Dynamics of electrons and explicit solutions of Dirac-Weyl systems}

\author{Alexander Sakhnovich}

\date{}
\begin{document}
\maketitle

\begin{abstract}   Explicit solutions of Dirac-Weyl system, which is essential in gra-phene
studies, are constructed using our recent approach to the construction of solutions of
dynamical systems. The obtained classes of solutions are much wider than the ones,
which were considered before. It is proved that neither constructed potentials nor 
corresponding solutions have singularities. Various examples are provided.

\end{abstract}



\section{Introduction}\label{Intro}
\setcounter{equation}{0}
It is well known that the motion of electron (in $\BR^2$)  in the presence of an electrostatic potential 
is often governed by the Dirac-Weyl system
\begin{align}& \label{I1}
\I \hbar v_F\left(\s_1 \frac{\p}{\p x}+\s_2 \frac{\p}{\p y}\right)\psi=\big(U(x,y)-E\big)\psi,
\end{align}
where $\hbar$ is the Planck constant, $v_F$ is the Fermi velocity, $U=\ov{U}$ and $E\in \BR$. 
Here $\ov{U}$ stands for the complex conjugate of $U$ and $\BR$ is the real axis. The matrices $\s_1$, $\s_2$ and $\s_3$ are
Pauli matrices:
\begin{align}& \label{I2}
\s_1=\begin{bmatrix} 0 & 1\\  1 & 0 \end{bmatrix}, \quad
\s_2=\begin{bmatrix} 0 & - \I\\  \I & 0 \end{bmatrix}, \quad \s_3=\begin{bmatrix} 1 & 0 \\  0 & -1 \end{bmatrix}.
\end{align}

Last years, the interest in graphene has essentially stimulated the study of this equation (see, e.g., \cite{HaP, HoRoy, Mid, SchUm, Stau}). In particular,
the important case when the scalar potential $U$ does not depend on the variable $y$ was recently studied,
for instance, in \cite{HaP, HoRoy, Mid} (see also some references therein). Assuming that $U$ does not depend on $y$ and multiplying both parts of \eqref{I1}
by $\frac{1}{\I \hbar v_F} \s_1$, we rewrite \eqref{I1} in an equivalent form
\begin{align}& \label{I3}
\psi_x=\I \s_3(-\psi_y+\I u(x)\s_2\psi), \quad \psi_x:=\frac{\p}{\p x}\psi, \quad u=\frac{E-U}{\hbar v_F}=\ov{u}.
\end{align}
Explicit solutions of \eqref{I3} are essential in the theory and applications. Some potentials generating explicit solutions
are treated  in \cite{HaP, HoRoy, Mid} after the separation of variables $\psi(x,y)=\E^{\I ky} \breve \psi(x)$ transforms
\eqref{I3} into the system depending on one variable. Well-known commutation methods \cite{D, Ge, GeT, KoSaTe, T} and several versions
 of B\"acklund Darboux  transformation (see, e.g., \cite{Ci, Gu, MS, SaA2, SaSaR, ZM}) could be fruitfully used in this case to produce explicit solutions of 
 linear systems 
 depending on one variable. However, a more complicated than $\psi(x,y)=\E^{\I ky} \breve \psi(x)$ dependence of the solutions $\psi$ on $y$ is of interest, and so we will apply
 some generalizations of our GBDT version (see \cite{GKS1, SaA2, SaA6, SaSaR} and references therein) of B\"acklund-Darboux transformation.
 Such generalizations (for the cases of linear
 systems of several variables) are given, for instance, in the papers \cite{FKRS, SaAJPhA2,  ALS-DCDS}. In particular, explicit solutions of non-stationary
 Dirac systems are constructed in \cite{FKRS}. Yet, taking into account that $u$ in \eqref{I3} does not depend on $y$, it seems more useful to
 modify here our approach to dynamical
 systems formulated in \cite{ALS-DCDS, SaA2016}.
 
 In Section \ref{GBDT} we present GBDT for  system \eqref{D2} which is somewhat more general than the Dirac-Weyl system \eqref{I3}. We give simple
 conditions under which the constructed (GBDT-transformed) potentials and solutions do not have singularities. Then, in Section \ref{Expl} we develop
a procedure to construct explicit solutions of systems \eqref{D2} and \eqref{I3} and consider various examples.
 
As usual, $\BN$ is the set of natural numbers, $\BR$ stands for the set of real numbers, $\BC$ stands for the set of complex numbers and
$\BC_+$ is the open upper half-plane. The notation $\ov{a}$ stands for the number  which is the complex conjugate of $a$.
The notation $\im(A)$ stands for the image of the matrix $A$, $A^*$ stands for the matrix  which is the complex conjugate transpose of $A$ and $I_m$ is the $m \times m$ identity matrix.
\section{GBDT for Dirac systems}\label{GBDT}
\setcounter{equation}{0}
In \cite[Section 7]{SaA2016} we constructed GBDT (generalized B\"acklund-Darboux transformation) for the dynamical system
\begin{align}& \label{D1}
z_x=J\big(-H_1(x)z_t+H_0(x)z\big),\quad J=-J^*, \quad H_1=H_1^*, \quad H_0=H_0^*.
\end{align}
Systems \eqref{D1} include dynamical Dirac systems
as a particular case. Here, we substitute into \eqref{D1} the variables $\psi$ and $y$ instead of $z$ and $t$, respectively. Since the equality $H_0=-H_0^*$ holds in our case,
we refuse from the requirement $H_0=H_0^*$ in \eqref{D1}. 
This change  is essential in further considerations. We set $J=\I \s_3$ and
$H_1=I_2$. Although the equality $H_0(x)=\I u(x) \s_2 $ $(u=\ov{u})$ corresponds to the case \eqref{I3}, in this section we do not require $u=\ov{u}$ and consider a more general
than \eqref{I3} system. Namely, we consider the system
\begin{align}& \label{D2}
\psi_x=\I \s_3(-\psi_y+ V(x)\psi), \quad V(x)=\begin{bmatrix} 0 & u(x)\\  -\ov{u(x)} & 0 \end{bmatrix}.
\end{align}
Further assume that $2\times 2$ matrix functions $Q_1(x)$ and $Q_0(x)$ are given and are locally summable on some interval $\cli$ such that
$0\in \cli$. (In our case $Q_1$ and $Q_2$ are the matrix coefficients in \eqref{D2}. More precisely, we have $Q_1\equiv -\I \s_3$ and $Q_0(x)=-\I\s_3 V(x)$.)
The following auxiliary result used also in  \cite{SaA2016} (see \cite[f-la (2.10)]{SaA2016}) is a particular case of much more general
GBDT-formulas \cite{SaA6, SaSaR}. We note that GBDT is based on the method of operator identities (see relations \eqref{D3} and  \eqref{D17}, and various references in \cite{SaSaR, SaL3}).
\begin{Pn} \label{PnAux} Fix $n\in \BN$ and five parameter matrices. Namely,  fix three $n\times n$ matrices $A_1$, $A_2$ and $S(0)$, and two $n\times 2$ matrices 
$\Pi_1(0)$ and $\Pi_2(0)$ such that the equality
\begin{align}& \label{D3}
A_1S(0)-S(0)A_2=\Pi_1(0)\Pi_2(0)^*
\end{align}
holds. For  $x\in \cli$, introduce matrix functions $\Pi_1(x)$, $\Pi_2(x)$ and $S(x)$ via their values at $x=0$ $($i.e., via the values $\Pi_1(0)$, $\Pi_2(0)$ and $S(0))$
and via equations
\begin{align}& \label{D4}
\Pi_1^{\prime}=A_1\Pi_1 Q_1+\Pi_1 Q_0, \quad (\Pi_2)^{\prime}=-A_2^*\Pi_2 Q_1^*-\Pi_2 Q_0^*, \quad S^{\prime}=\Pi_1 Q_1 \Pi_2^*,
\end{align}
where $\Pi_1^{\prime}(x)=\big(\frac{d}{dx} \Pi_1\big)(x)$. 

Then, in the points of invertibility of $S(x)$, we have
\begin{align}& \label{D5}
\big(\Pi_2^*S^{-1}\big)^{\prime}=-Q_1\Pi_2^*S^{-1}A_1-\wt Q_0\Pi_2^*S^{-1},
\end{align}
where
\begin{align}& \label{D6}
 \wt Q_0:=Q_0- \big(Q_1X-XQ_1\big), \quad X:=\Pi_2^*S^{-1}\Pi_1.
\end{align}
\end{Pn}
\begin{Rk}\label{RkInv1} Further, speaking about $S(x)^{-1}$, we mean $S(x)^{-1}$ in the points where
$S(x)$ is invertible. The invertibility of $S(x)$ is discussed, in particular, in Proposition \ref{PnInv}
and in Remark \ref{RkOpId}.
\end{Rk}
In our case, that is, in the case of system \eqref{D2} we set
\begin{align}& \label{D7}
 A=A_1=A_2^*, \quad S(0)=S(0)^*, \quad \Pi(0)=\Pi_1(0)=\I \Pi_2(0),
\end{align}
so that \eqref{D3} takes the form
\begin{align}& \label{D8}
AS(0)-S(0)A^*=\I \Pi(0)\Pi(0)^* \qquad \big(S(0)=S(0)^*\big).
\end{align}
As mentioned above we put
\begin{align}& \label{D9}
Q_1\equiv -\I \s_3, \quad Q_0(x)=-\I\s_3 V(x),
\end{align}
which yields $Q_i^*=-Q_i$ ($i=0,1$). Therefore, since $A=A_1=A_2^*$, equations on $\Pi_1$ and $\Pi_2$ in \eqref{D4} coincide
and taking into account $\Pi(0)=\Pi_1(0)=\I \Pi_2(0)$ we can introduce $\Pi(x)$ by the equalities
$\Pi(x)=\Pi_1(x)=\I \Pi_2(x)$. Next, we rewrite formulas \eqref{D4}--\eqref{D6} (from Proposition \ref{PnAux}) in the form
\begin{align}& \label{D10}
\Pi^{\prime}=A\Pi  Q_1+\Pi Q_0, \quad  S^{\prime}=\I \Pi Q_1 \Pi^*;
\\  & \label{D11}
\big(\Pi^*S^{-1}\big)^{\prime}=-Q_1\Pi^*S^{-1}A-\wt Q_0\Pi^*S^{-1};
\\ & \label{D12}
 \wt Q_0:=Q_0- \I \big(Q_1\clx -\clx Q_1\big), \quad \clx :=\Pi^*S^{-1}\Pi.
\end{align}
Since $S(0)=S(0)^*$ and $S^{\prime}(x)=S^{\prime}(x)^*$ we have 
\begin{align}& \label{D12'}
S(x)=S(x)^*, \quad \clx(x)=\clx(x)^*.
\end{align}
Recall that $A$ and $S(0)$ are $n \times n$ matrices and $\Pi(0)$ is an $n \times 2$ matrix.
GBDT-transformed system of the form \eqref{D2} is determined by the initial system \eqref{D2}
and a triple of  matrices $\{A, S(0), \Pi(0)\}$ such that \eqref{D8} holds.
More precisely,  relations \eqref{D10}--\eqref{D12} imply our next statement.
\begin{Tm} \label{TmGBDT} Let the initial system \eqref{D2} and a triple of  matrices \\ $\{A, S(0), \Pi(0)\}$ $($such that \eqref{D8} holds$)$
be given. Then the vector functions
\begin{align}& \label{D13}
\wt \psi(x,y)=\Pi(x)^*S(x)^{-1}\E^{- y A}h, \quad h\in \BC^n,
\end{align}
where $\Pi(x)$ and $S(x)$ are defined by $\Pi(0)$, $S(0)$ and relations \eqref{D10} and \eqref{D9}, satisfy
the GBDT transformed system
\begin{align}& \label{D2'}
\wt \psi_x=\I \s_3(-\wt \psi_y+ \wt V(x)\wt \psi), \quad \wt V:=V+\I(\s_3\clx \s_3-\clx).
\end{align}
Here $\clx=\Pi^*S^{-1}\Pi$ and $\wt V$ has the same structure as $V$, that is, $ \wt V=\begin{bmatrix} 0 & \wt u\\  -\ov{\wt u} & 0 \end{bmatrix}$.
\end{Tm}
\begin{proof}. The structure of $\wt V$ follows from the second equalities in \eqref{D2}, \eqref{D12}, \eqref{D12'} and \eqref{D2'}.
Moreover, in view of \eqref{D9}, \eqref{D12} and the second equality in \eqref{D2'}, we have
\begin{align}& \label{D14}
\I \s_3 \wt V=-\wt Q_0 \quad {\mathrm{and}} \quad  \I \s_3 \wt V\wt \psi =-\wt Q_0 \wt \psi.
\end{align}
Clearly, \eqref{D13} yields
\begin{align}& \label{D15}
-\I \s_3 \wt \psi_y= -Q_1\Pi(x)^*S(x)^{-1}A\E^{- y A}h.
\end{align}
The terms of the right-hand side of system \eqref{D2'} are rewritten in \eqref{D14} and \eqref{D15}. Now, using \eqref{D11} and \eqref{D13} we see that the
 left-hand side of the system \eqref{D2'} coincides with its right-hand side, that is, $\wt \psi$ given by \eqref{D13} satisfies the system \eqref{D2'}.
\end{proof}
Equality \eqref{D8} and equations \eqref{D10} yield the identity
\begin{align}& \label{D16}
AS(x)-S(x)A^*=\I \Pi(x)\Pi(x)^* \quad (x\in \cli).
\end{align}
Indeed, in view of $Q_i^*=-Q_i$, direct differentiation shows that  the derivatives of  both parts of \eqref{D16}
coincide for all $x \in \cli$, and according to \eqref{D8} the identity \eqref{D16} is valid at $x=0$.
Thus, \eqref{D16} holds for all $x \in \cli$.

The next proposition  on the invertibility
of $S$ and on $\s(A)$ (where $\s(A)$ denotes the spectrum of $A$) easily follows from \eqref{D16}. 
\begin{Pn} \label{PnInv} Let  the inequality $S(0)>0$ and the conditions of Theorem \ref{TmGBDT}  hold.  
Then, we have $S(x)>0$ $($and so $S(x)$ is invertible$)$ for all $x \in \cli$.
Moreover, we have $\s(A)\in (\BC_+ \cup \BR)$. 
\end{Pn}
\begin{proof}.  The second equality in \eqref{D10} and the identity \eqref{D16} imply inequalities
\begin{align}\nn 
\big(\E^{-\I x A}S(x)\E^{\I x A^*}\big)^{\prime}&=\E^{-\I x A}\big(S^{\prime}(x)-\I (AS(x)-S(x)A^*)\big)\E^{\I x A^*}
\\ \label{Dd1} &
=\E^{-\I x A}\Pi(x)\big(\s_3+ I_2\big)\Pi(x)^*\E^{\I x A^*}\geq 0;
\\ \nn
\big(\E^{\I x A}S(x)\E^{-\I x A^*}\big)^{\prime}&=\E^{\I x A}\big(S^{\prime}(x)+\I (AS(x)-S(x)A^*)\big)\E^{-\I x A^*}
\\ \label{Dd2} &
=\E^{\I x A}\Pi(x)\big(\s_3- I_2\big)\Pi(x)^*\E^{-\I x A^*}\leq 0.
\end{align}
By virtue of \eqref{Dd1} we derive
\begin{align}& \label{Dd3}
\E^{-\I x A}S(x)\E^{\I x A^*} \geq S(0)>0 \quad {\mathrm{for}} \quad x\geq 0.
\end{align}
By virtue of \eqref{Dd2} we derive
\begin{align}& \label{Dd4}
\E^{\I x A}S(x)\E^{-\I x A^*} \geq S(0)>0 \quad {\mathrm{for}} \quad x\leq 0.
\end{align}
It is immediate (from \eqref{Dd3} and \eqref{Dd4})  that $S(x)>0$ everywhere on $\cli$. 

Rewriting \eqref{D16} at $x=0$ as 
$$S(0)^{-1/2}AS(0)^{1/2}- \big(S(0)^{-1/2}AS(0)^{1/2}\big)^*=\I S(0)^{-1/2}\Pi(0)\Pi(0)^*S(0)^{-1/2},$$
we obtain the relation  $\s(A)\in (\BC_+ \cup \BR)$.
\end{proof}

\begin{Rk} \label{RkOpId} We note $($see \cite{SaA2, SaA6, SaSaR}$)$ that identities
of the form \eqref{D16} appear in GBDT for skew-selfadjoint spectral Dirac systems whereas identities
\begin{align}& \label{D17}
AS(x)-S(x)A^*=\I \Pi(x)\s_3\Pi(x)^* \quad (x\in \cli). 
\end{align}
 appear in the selfadjoint case. For nonlinear integrable equations
with auxiliary linear Dirac  or, equivalently, ZS-AKNS systems $($e.g., for nonlinear Schr\"odinger and mKdV
equations$)$ the identity \eqref{D16} appears in the case of multisoliton-type potentials whereas the identity
\eqref{D17} corresponds to potentials with singularities \cite{GKS1, FKS, SaA8}.
Some discussion on multisoliton-type potentials and connections between system \eqref{I1} and mKdV
is contained also in \cite{HoRoy}.
\end{Rk}

\section{Explicit solutions of Dirac-Weyl systems}\label{Expl}
\setcounter{equation}{0}
Explicit solutions  usually appear in B\"acklund-Darboux transformations when we choose trivial initial systems.
Indeed, setting in \eqref{D2} $V \equiv 0$ we rewrite the equation on $\Pi$ in \eqref{D10} as
$\Pi^{\prime}=-\I A\Pi \s_3$, which yields
\begin{align}& \label{Ex1}
\Pi(x)=\begin{bmatrix} \Lam_1(x) & \Lam_2(x) \end{bmatrix}, \quad \Lam_1(x)=\E^{-\I x A}\Lam_1(0), 
\quad \Lam_2(x)=\E^{\I x A}\Lam_2(0),
\end{align}
where $\Lam_k$  $(k=1,2)$ are the columns of $\Pi$. Explicit expressions for $\Pi$
provide explicit expressions for $S$, $\wt V$ and $\wt \psi$. In particular, we have
\begin{align}& \label{Ex2}
S(x)=S(0)+\int_0^x
\Pi(r)\s_3\Pi(r)^*dr \quad(x>0), \\
& \label{Ex3}
 S(x)=S(0)-\int_x^0
\Pi(r)\s_3\Pi(r)^*dr \quad(x<0)
\end{align}
with $\Pi$ determined in \eqref{Ex1}. The corresponding explicit solutions $\wt \psi$ are given in the next corollary
of Theorem \ref{TmGBDT}.
\begin{Cy} \label{MCy} Let  some triple of  parameter matrices $\{A, S(0)>0, \Pi(0)\}$ $($such that \eqref{D8} holds$)$
be fixed. Introduce matrix functions $\Pi(x)$ and $S(x)$ by the relations \eqref{Ex1}--\eqref{Ex3}.
Then, the vector functions $\wt \psi$ of the form \eqref{D13} satisfy $($are explicit solutions of $)$
the Dirac system
\begin{align}& \label{Ex4}
\wt \psi_x=\I \s_3(-\wt \psi_y+ \wt V(x)\wt \psi); \quad \wt V:=\I(\s_3\clx \s_3-\clx), \quad \clx=\Pi^*S^{-1}\Pi
\end{align}
on the real axis $-\infty<x<\infty$. The matrix function $S(x)$ in \eqref{D13} and \eqref{Ex4} is positive definite $($i.e., $S(x)>0)$, and so invertible,
everywhere on $\BR$. Thus, $\wt V$ and $\wt \psi$ are well-defined  everywhere on $\BR$.

Under additional condition that the entries of the  matrices $\I A$ and $S(0)$ and of the vectors $\I \Lam_1(0)$ and $\Lam_2(0)$ are all real-valued,
the potential $\wt V$ is real-valued as well, and we can rewrite \eqref{Ex4} in the form  of the Dirac-Weyl system \eqref{I3}$:$
\begin{align}& \label{Ex5}
\wt \psi_x=\I \s_3(-\wt \psi_y+ \I \wt u(x)\s_2 \wt \psi); \quad \wt u:=-2\I\Lam_1^* S^{-1}\Lam_2=\ov{\wt u}.
\end{align}
\end{Cy}
\begin{proof}. The fact that $\wt \psi$ given by \eqref{D13} satisfies \eqref{Ex4} is immediate from Theorem \ref{TmGBDT}.
The positive definiteness of $S(x)$ follows from Proposition \ref{PnInv}. Finally, additional conditions in  Corollary \ref{MCy}
and relations \eqref{Ex1}--\eqref{Ex3} imply that the entries of $\I \Lam_1(x)$, $\Lam_2(x)$ and $S(x)$ are real-valued.
Hence, $\wt u$ given by the second equality in \eqref{Ex5} is real-valued. For $\wt V$ introduced in \eqref{Ex4}
and the real-valued $\wt u$ from \eqref{Ex5}, we have the equalities
\begin{align}& \label{Ex6}
V=\begin{bmatrix} 0 & \wt u \\ - \wt u \end{bmatrix}= \I \wt u \s_2 ,
\end{align}
and substituting \eqref{Ex6} into the Dirac  system  \eqref{Ex4} we obtain the Dirac-Weyl system \eqref{Ex5}.
\end{proof}
Below we give several examples (and classes of examples) of the triples $\{A, S(0), \Pi(0)\}$ of parameter 
matrices which satisfy all the conditions of Corollary \ref{MCy}. For simplicity, we set $S(0)=I_n$ in these examples.
We set $A=\I \cla$, where $\cla$ is a matrix with real-valued entries.
GBDT allows to consider cases of non-diagonal matrices $A$ (in which situation explicit solutions with interesting properties appear; see, e.g.,
\cite{Abl, SaAx}), and in Example \ref{Ee2} we calculate
$\wt u$ and $\wt \psi$ for the case when $A$ is similar to a $2\times 2$ Jordan cell.
\begin{Ee}\label{Ee1} If $A=\I \cla$ is a scalar and $S(0)=1$  $($i.e.,  the order $n$ of $A$ and $S$ equals $1)$,
then $\Lam_1(0)$ and $\Lam_2(0)$ are scalars as well and the identity \eqref{D8} takes the form
$2 \cla =|\Lam_1(0)|^2+|\Lam_2(0)|^2$. All values $\cla$ and $|\Lam_k(0)|$ $(k=1,2)$, which satisfy this equality,
generate corresponding parameters $A=\I \cla$ and $\Pi(0)= \begin{bmatrix}\pm \I |\Lam_1(0)| & \pm |\Lam_2(0)|\end{bmatrix}$
or $\Pi(0)= \begin{bmatrix}\pm \I |\Lam_1(0)| & \mp |\Lam_2(0)|\end{bmatrix}$ satisfying conditions of
Corollary \ref{MCy}.
\end{Ee}
\begin{Ee}\label{Ee2} In the case $n=2$, we consider the example$:$ $S(0)=I_2$,
\begin{align}& \label{Ex7}
A=\I \cla=\I \begin{bmatrix} 1 & 0 \\ 1 & 1 \end{bmatrix},  \quad 
\Pi(0)= \begin{bmatrix} \Lam_1(0) & \Lam_2(0) \end{bmatrix}=\frac{1}{\sqrt{2}}\begin{bmatrix} 2 \I & 0 \\ \I & \sqrt{3} \end{bmatrix}.
\end{align}
Clearly, \eqref{D8} holds for this example. Simple calculations show that
\begin{align}& \label{Ex8}
\E^{\mp \I x A}=\E^{\pm x \cla}=\E^{\pm x}\E^{\pm x(\cla -I_2)}=\E^{\pm x}
\begin{bmatrix} 1 & 0 \\  \pm x & 1 \end{bmatrix}.
\end{align}
Using \eqref{Ex1}, \eqref{Ex7} and \eqref{Ex8} we obtain
\begin{align}& \label{Ex9}
\Lam_1(x)=\frac{1}{\sqrt{2}} \E^{x}
\begin{bmatrix} 2 \I  \\  \I(2 x +1) \end{bmatrix}, \quad \Lam_2(x)=\frac{1}{\sqrt{2}} \E^{-x}
\begin{bmatrix} 0  \\  \sqrt{3} \end{bmatrix};
\\ & \label{Ex10}
\Pi(x)\s_3\Pi(x)^*=\frac{1}{2}\left(\E^{2x}\begin{bmatrix} 2  \\  2x+1 \end{bmatrix}
\begin{bmatrix} 2  & 2x+1 \end{bmatrix}
-3\E^{-2x}
\begin{bmatrix} 0 & 0 \\ 0 &  1 \end{bmatrix}\right).
\end{align}
Integrating \eqref{Ex10} and substituting the result into \eqref{Ex2} and \eqref{Ex3}, we
calculate $S$ and easily derive from the expression for $S(x)$ that  $\det S(x)=\frac{1}{4}\big(\E^{4x}+3\big)$ and
\begin{align}& \label{Ex11}
 S(x)^{-1}=\frac{4}{\E^{4x}+3}
\begin{bmatrix}
\frac{1}{4}\big((4x^2+1)\E^{2x}+3\E^{-2x}\big) & -x\E^{2x} \\
-x \E^{2x} & \E^{2x}
 \end{bmatrix}.
\end{align}
It is immediate from \eqref{Ex5}, \eqref{Ex9} and \eqref{Ex11} that
\begin{align}& \label{Ex12}
\wt u(x) =-2\I \Lam_1(x)^*S(x)^{-1}\Lam_2(x)=-\frac{4\sqrt{3}\, \E^{2x}}{\E^{4x}+3}.
\end{align}
Next, we note that
\begin{align}& \label{Ex13}
\E^{-yA}=\E^{-\I y}\E^{-\I y(\cla -I_2)}=\E^{-\I y}\begin{bmatrix} 1 & 0 \\ - \I y &  1 \end{bmatrix}.
\end{align}
Finally, relations \eqref{Ex7}, \eqref{Ex9}, \eqref{Ex11} and \eqref{Ex13} yield the equalities$:$
\begin{align} \nn
\Pi(x)^* S(x)^{-1}\E^{-yA}h=&\frac{2 \sqrt{2}\, \E^{-\I y}}{\E^{4x}+3}
\begin{bmatrix}
\frac{\I}{2}\big((2x-1)\E^{3x}-3\E^{-x}\big) & -\I \E^{3x} \\
-\sqrt{3} \, x \E^{x} & \sqrt{3}\, \E^{x}
 \end{bmatrix}
\\ & \label{Ex14} \times  
 \begin{bmatrix} 1 & 0 \\ - \I y &  1 \end{bmatrix}h, \quad h \in \BC^2.
\end{align}
Thus, we obtain the following corollary.
\end{Ee}
\begin{Cy} \label{CyEx} Vector functions $\wt \psi$ of the form \eqref{Ex14}
are solutions of the Dirac-Weyl system \eqref{Ex5} where $\wt u$ is given by \eqref{Ex12}.
\end{Cy}
Now, we introduce wide classes of parameter triples $\{A, S(0), \Pi(0)\}$ (satisfying
the conditions of Corollary \ref{MCy}) with  an arbitrary
fixed number $n\in \BN$. As before, we assume that $S(0)=I_n$ and $A=\I \cla$,
where the entries of the $n\times n$ matrix $\cla$ are real-valued.
\begin{Ee}\label{Ee3} Set
\begin{align}& \label{Ex15}
\cla=\cla_0+h_1h_1^*+h_2 h_2^* \quad (\cla_0^*=-\cla_0), \quad  \Pi(0)=\sqrt{2} \begin{bmatrix} \I h_1 & h_2 \end{bmatrix},
\end{align}
where  $h_k\in \BR^n$ for $k=1,2$, and the entries of $\cla_0$ are real-valued. Then all the conditions of Corollary \ref{MCy}
are fulfilled.
\end{Ee}
\begin{Ee}\label{Ee4} Set $\Pi(0)= \begin{bmatrix} \I h_1 & h_2 \end{bmatrix}$, where  $h_k\in \BR^n$ for $k=1,2$, and choose
a triangular matrix $\cla$ such that $\cla+\cla^*=\Pi(0)\Pi(0)^*$.
 Then all the conditions of Corollary \ref{MCy}
are fulfilled.
\end{Ee}

\bigskip

\noindent{\bf Acknowledgments.}
 {This research   was supported by the
Austrian Science Fund (FWF) under Grant  No. P29177.}

\begin{flushright}

A.L. Sakhnovich,\\
Fakult\"at f\"ur Mathematik, Universit\"at Wien, \\
Oskar-Morgenstern-Platz 1, A-1090 Vienna, Austria\\
e-mail: oleksandr.sakhnovych@univie.ac.at
\end{flushright}



\begin{thebibliography}{AGKS}
\bibitem{Abl}
 Ablowitz M J, Chakravarty S, Trubatch A D and Villarroel J 2000 A novel class of solutions of the non-stationary
Schr\"odinger and the KadomtsevÐPetviashvili I equations 
\textit{Phys. Lett. A} \textbf{267} 132--146

\bibitem{Ci}
Cieslinski J L 2009
Algebraic construction of the Darboux matrix revisited 
 \textit{J. Phys. A} \textbf{42} 404003

\bibitem{D}
Deift P A 1978 Applications of a commutation formula \textit{Duke Math. J.} \textbf{45}  267--310

\bibitem{FKRS}
Fritzsche B, Kirstein B, Roitberg I and Sakhnovich A L 2015
Pseudo-exponential-type solutions of wave equations depending on several variables 
\textit{SIGMA} \textbf{11} 010

\bibitem{FKS}
Fritzsche B, Kirstein B and Sakhnovich A L 2006 
Completion problems and scattering problems for Dirac type differential equations with singularities  \textit{J. Math. Anal. Appl.} \textbf{317} 510--525 

 \bibitem{Ge}
Gesztesy F 1993
A complete spectral characterization of the double commutation method  \textit{J. Funct. Anal.}  \textbf{117}  401--446

\bibitem{GeT}
Gesztesy F and Teschl G 1996 
On the double commutation method \textit{Proc. Amer. Math. Soc.} \textbf{124} 1831--1840

\bibitem{GKS1}
Gohberg I, Kaashoek  M A and Sakhnovich A L 1998 Canonical systems with rational spectral densities: explicit formulas and applications \textit{ Math. Nachr.} \textbf{194} 93--125

\bibitem{Gu}
Gu C, Hu H and  Zhou Z 2005  \textit{Darboux transformations in integrable systems. Theory and their applications to geometry} Mathematical Physics Studies \textbf{26} (Dordrecht: Springer)

\bibitem{HaP}
Hartmann R R and Portnoi M E 2014 Quasi-exact solution to the Dirac equation for the hyperbolic-secant potential \textit{Phys. Rev. A} \textbf{89}  012101

\bibitem{HoRoy}
Ho C-L and Roy P 2015 mKdV equation approach to zero energy states of graphene \textit{Europhysics Letters} \textbf{112} 47004

\bibitem{KoSaTe}
Kostenko A, Sakhnovich A and  Teschl G 2012
Commutation methods for Schr\"odinger operators with strongly singular potentials \textit{Math. Nachr.} \textbf{285} 392--410

\bibitem{MS}
Matveev V B and Salle M A 1991 \textit{Darboux transformations and
solitons} (Berlin: Springer Verlag)

\bibitem{Mid}
Midya B and Fernandez D J 2014 Dirac electron in graphene under supersymmetry generated magnetic fields \textit{J. Phys. A} \textbf{47}  285302

\bibitem{SaA2}
Sakhnovich A L  1994 
 Dressing procedure for solutions of nonlinear equations and the method of operator identities \textit{Inverse Problems}  \textbf{10}  699--710

\bibitem{SaA6} Sakhnovich A L 2001 
Generalized B\"acklund--Darboux transformation: spectral properties and nonlinear equations \textit{J. Math. Anal. Appl.} \textbf{262}  274--306

 \bibitem{SaAx}
 Sakhnovich A L 2003 Matrix Kadomtsev--Petviashvili equation: matrix identities and explicit non-singular solutions  \textit{J. Phys. A} \textbf{36}  5023--5033

\bibitem{SaA8}
Sakhnovich A L  2003  Dirac type system on the axis: explicit
formulas for matrix potentials with singularities and soliton-positon interactions \textit{Inverse Problems} \textbf{19} 845--854 

  \bibitem{SaAJPhA2}
 Sakhnovich A L 2011 The time-dependent Schr\"odinger equation of dimension $k+1$: explicit and rational solutions via GBDT and multinodes \textit{J. Phys. A} \textbf{44}  475201
  
 \bibitem{ALS-DCDS}
Sakhnovich A L 2016 Dynamical canonical systems and their explicit solutions \textit{arXiv:1603.08709} Discrete Contin. Dyn. Syst. to appear

\bibitem{SaA2016}
Sakhnovich A L  2016
Hamiltonian systems and Sturm--Liouville equations: Darboux transformation and applications \textit{arXiv:1608.02348}
  

\bibitem{SaSaR}
Sakhnovich A L, Sakhnovich L A   and Roitberg I 2013  \textit{Inverse Problems and Nonlinear Evolution Equations. 
 Solutions, Darboux Matrices and Weyl--Titchmarsh Functions} {De Gruyter Studies in Mathematics} \textbf{47}  (Berlin: De Gruyter)


\bibitem{SaL3}
Sakhnovich L A 1999
\textit{Spectral theory of canonical differential
systems, method of operator identities}  Operator Theory Adv.
Appl. \textbf{107} (Basel: Birkh{\"a}user)

\bibitem{SchUm}
 Schmidt K M and  Umeda T  2015 Schnol's theorem and spectral properties of massless Dirac operators with scalar potentials \textit{Lett. Math. Phys.} \textbf{105}  1479--1497

\bibitem{Stau} 
Stauber T 2014 
Plasmonics in Dirac systems: from graphene to topological insulators 
 \textit{J. Phys.: Condens. Matter} \textbf{26} 123201
 
\bibitem{T} 
 Teschl G 1998 Deforming the point spectra of one-dimensional Dirac operators \textit{Proc. Amer. Math. Soc.}  \textbf{126} 2873--2881 
 
\bibitem{ZM}
Zakharov V E and Mikhailov A V 1980 On the integrability of
classical spinor models in two-dimensional space-time \textit{Comm.
Math. Phys.} \textbf{74}  21--40
 
\end{thebibliography}
\end{document}